\documentclass[a4paper,fleqn,usenatbib]{mnras}

\usepackage{newtxtext,newtxmath}

\usepackage[T1]{fontenc}
\usepackage{ae,aecompl}

\usepackage{amssymb} 
\usepackage{amsmath} 
\usepackage{hyperref}
\usepackage{booktabs,longtable}
\usepackage{graphicx} 
\usepackage{wasysym} 
\usepackage{lscape} 
\usepackage{caption} 
\usepackage{wrapfig,lipsum,booktabs} 
\usepackage{enumitem} 
\usepackage{mathtools} 
\usepackage{overpic} 
\usepackage{graphbox} 
\usepackage{tabularx} 
\usepackage{caption} 
\usepackage[table]{xcolor} 
\usepackage{multirow} 

\newcommand{\GALFIT}{\textsc{Galfit}}
\newcommand{\HST}{\emph{HST}}
\newcommand{\SKIRT}{\textsc{Skirt}}
\newcommand{\Sersic}{S\'{e}rsic}

\newcommand{\Chisq}{$\chi_{\nu}^{2}$}

\newlist{cutenumerate}{enumerate}{1}
\setlist[cutenumerate,1]{
  label={\arabic*},
  leftmargin=*,
  align=left,
  labelsep=0.05cm,
}

\title
[Modelling dust rings] 
{
 Modelling dust rings in early-type galaxies through a sequence of
 radiative transfer simulations and 2D image fitting
} 

\author[P. Bonfini et al.]
{
 \parbox{\textwidth}{
  P. Bonfini$^{1}$\thanks{E-mail: \texttt{p.bonfini@irya.unam.mx}},
  O. Gonz{\'a}lez-Mart{\'i}n$^{1}$,
  J. Fritz$^{1}$,
  T. Bitsakis$^{1}$,
  G. Bruzual$^{1}$
  and
  B. Cervantes Sodi$^{1}$
 }
 \vspace{0.4cm}\\
 \parbox{\textwidth}{
  $^{1}$Instituto de Radioastronom{\'i}a y Astrof{\'i}sica,
  Universidad Nacional Aut{\'o}noma de M{\'e}xico, Morelia 58089,
  M{\'e}xico.\\
 }
}

\date{Accepted XXX. Received YYY; in original form ZZZ}

\pubyear{2017}

\begin{document}
\label{firstpage}
\pagerange{\pageref{firstpage}--\pageref{lastpage}}
\maketitle

\begin{abstract}
 A large fraction of early-type galaxies (ETGs) host prominent dust features,
 and central dust rings are arguably the most interesting among them.
 We present here `Lord Of The Rings' (LOTR), a new methodology which allows to
 integrate the extinction by dust rings in a 2D fitting modelling of the
 surface brightness distribution.
 Our pipeline acts in two steps, first using the surface fitting software \GALFIT{}
 to determine the unabsorbed stellar emission, and then adopting the radiative
 transfer code \SKIRT{} to apply dust extinction.
 We apply our technique to NGC~4552 and NGC~4494, two nearby ETGs.
 We show that the extinction by a dust ring can mimic, in a surface brightness
 profile, a central point source (e.g.\ an unresolved nuclear stellar cluster or
 an active galactic nucleus; AGN) superimposed to a `core' (i.e.\ a central
 flattening of the stellar light commonly observed in massive ETGs).
 We discuss how properly accounting for dust features is of paramount importance
 to derive correct fluxes especially for low luminosity AGNs (LLAGNs).
 We suggest that the geometries of dust features are strictly connected with
 how relaxed is the gravitational potential, i.e.\ with the evolutionary stage of
 the host galaxy.
 Additionally, we find hints that the dust mass contained in the ring relates to
 the AGN activity. 
\end{abstract}

\begin{keywords}
  galaxies: structure ---
  galaxies: individual: NGC~4552 (M~89, UGC~7760, VCC~1632), NGC~4494 (UGC~7662) ---
  galaxies: evolution ---
  galaxies: nuclei ---
  galaxies: intergalactic medium
\end{keywords}


\section{Introduction}
\label{Introduction}

\noindent
Early-type galaxies (ETGs) are long known to host dust features of different
nature, ranging from extended lanes crossing the whole galaxy, to inner disk
or ring structures \citep[e.g.][]{rest,lauer:2005}.
These peculiarities are usually overlooked in the analysis aimed at performing
surface brightness fitting of the galaxy surface brightness distribution,
due to the complexity of including a parametric model accounting for dust
extinction.
Hence, the approach used in arguably all studies is to mask the dust-affected
area and exclude the corresponding data points from the fit.

Although sometimes unavoidable, this procedure is at least controversial, since
the range of data exclusion is arbitrary, and can therefore unpredictably affect
the choice of best-fit model.
The most questionable situations are those involving the presence of a central
dust ring/disk.
The absorption produced by these features can in fact be indistinguishable from
a real galaxy ``core'', i.e.\ an intrinsic light deficit due to a lack of stars,
common in massive galaxies \citep[$M_B$ < $-20.5$~mag; e.g.][]{kormendy:2009,graham:2016}.
All the studies regarding the central surface brightness of ETGs to date have
preferentially excluded problematic, dusty objects from their analysis.
However, in several cases, surface brightness fits have been attempted 
even in the presence of obvious obscuration (after heavy masking), and conclusions
were drawn over the core/core-\emph{less} nature of the objects
\citep[e.g.][]{lauer:2007,ferrarese:2006,richings,dullo:2014}.

On top of this issue, an additional complication regards the potential presence of
a central active galactic nucleus (AGN), and in particular of low-luminosity
AGNs (LLAGNs).
In fact, these objects posses an intrinsically low luminosity
\citep[$L_{bol} < 10^{42}$~erg/sec, $L_{bol} \sim 30 \times L_X$; e.g.][]{ho:AGN_review},
and hence they are easily over-shined by the high central intensity of
core-\emph{less} galaxies with large \Sersic{} indexes.
Core galaxies are instead the ideal hosts to look for such weak AGNs, because
their `flatter' inner profiles allow the LLAGN to stand out as a central point
source.
However, in optical/UV imaging, a central LLAGN superimposed
to a core is observationally undistinguishable from a classical core-\emph{less}
galaxy (i.e.\ a typical \Sersic{} galaxy) shining through the hole of a inner
dust ring.
In fact, for a sufficiently small ring hole and for an appropriate ring orientation,
the light passing through the hole will appear as a point-like source
(see Figure \ref{figure:fit_models}), while the dust absorption will mimic a stellar depleted core.
The observational evidence that the vast majority of LLAGNs are associated
to circumnuclear dust \citep{gonzalez-delgado} further complicates this picture.
Given that core galaxies are such an ideal benchmark to study those objects,
it is of paramount importance to disentangle any role of the dust in producing
central flattenings in the surface brightness distribution.

The study of dust components in galaxies have been significantly progressing
during the last two decades.
Not only the influence of dust  over the kinematics of both late- and early-type
galaxies has been ascertained \citep[e.g.][]{baes:2002,baes:2003}, but radiative transfer
codes have shown how dust biases the observed bulge/disk structural parameters
\citep[e.g.][]{gadotti:2010}.
In particular, radiative transfer modelling has been extensively applied to perform
surface brightness fitting of dust-obscured, edge on disk galaxies
\citep[e.g.][]{emsellem:1995,xilouris:1999,baes:2010,delooze:2012,viaene:2015} with the
aim of characterizing the galaxy's spectral energy distribution and hence
the source of dust heating, or addressing the so-called energy balance issue
\citep[e.g.][and references therein]{saftly:2015}.
Here  we intend to adopt a similar approach to solve the core/AGN/dust ambiguity in ETGs.
We propose a new, simple, but
powerful methodology to derive central dust geometries and masses.
The technique involves creating, using a radiative transfer code, a pool of
model images of the galaxy in which the effects of dust are taken into account.
These 3D models are projected on the field of view to produce the 2D surface
brightness models, which are then compared with the actual image.
In this way we can infer the bona-fide dust distribution.
One immediate application (among others) is to distinguish central point sources
into actual AGNs and stellar light shining through patches in the dust, based
uniquely on imaging.
In this paper, we provide one example of such application.
In Section \ref{Sample and data} we present the sample of dust-obscured galaxies
and the data chosen for this study; in Section \ref{Modelling of dust rings}
we present and apply our methodology, i.e.\ the radiative transfer/2D fitting of
dust features; in Section \ref{Discussion} we discuss our
results in the context of the identification of dust disks/rings, cores and LLAGN
from the morphological analysis of galaxies;
finally, in Section \ref{Summary and conclusions} we summarize our conclusions.

\section{Sample and data}
\label{Sample and data}

\begin{table*}
 \centering

 \begin{tabular*}{\textwidth}{l @{\extracolsep{\fill}} ccccccc}
  \hline
  \multicolumn{8}{c}{\textsc{Details of the sample images}} \\
  \hline
  \hline
  \addlinespace 
  \multicolumn{1}{c}{Target}     & RA (J2000) & Dec (J2000) & $D$        & Camera/Filter & Exp. time  & Prop. ID   & Reference \\
                                 & [hh:mm:ss] & [dd:mm:ss]  & [Mpc]      &               & [s]        &            &           \\
  \multicolumn{1}{c}{\tiny{(1)}} & \tiny{(2)} & \tiny{(3)}  & \tiny{(4)} & \tiny{(5)}    & \tiny{(6)} & \tiny{(7)} & \tiny{(8)}\\
  \hline
  \addlinespace 
NGC~4552 & 12:35:39.8 & +12:33:23 & 16.0 & ACS/$F475W$   & 750  & 9401 & \cite{ferrarese:2006} \\
NGC~4494 & 12:31:24.1 & +25:46:31 & 13.7 & WFPC2/$F555W$ & 1000 & 5454 & \cite{forbes:NGC4494}\\
  \hline

 \end{tabular*}
 \parbox{\textwidth}{
  \caption[]{
   Details of the HLA products used in this work.
   These are sky-subtracted mosaics composed by same-orbit observations
   for each galaxy.
   \\
   (1) Target name.
   (2,3) Target coordinates from NED (J2000).
   (4) Target distance from NED.
   (5) \HST{} camera and filter.
   (6) Exposure time.
   (7) Proposal ID for the archival observation.
   (8) First publication for the data set.
   \label{table:sample_images} 
  }
 }
\end{table*}

\noindent
The test galaxies for our study are NGC~4552 (M~89, UGC~7760, VCC~1632; Virgo) and
NGC~4494 (UGC~7662; Coma).
Deep \HST{} images clearly revealed that these objects host central disk- or ring-like
dust structures (see Figure \ref{figure:fit_models}).
In particular, NGC~4552 hosts a quasi face-on disk surrounded by several dust
streams which spread radially beyond the effective radius of the galaxy
\citep[][]{ferrarese:2006}. 
The structure in NGC~4494, first observed by \cite{forbes:NGC4494}, resembles
instead a more edge-on ring, or better a torus with a large inner hole and a
small opening angle.
These two objects have been chosen from the ensemble of dust rich ETGs of
\citet[][their Figure 1]{lauer:2005} in order to be representative of the
two classes of core (NGC~4552) and core-\emph{less} (NGC~4494) galaxies,
according to the central classification provided by the same authors.
In particular, the core size of NGC~4552 was estimated in 0.49$\arcsec$
(42.5~pc).

An additional criterion for the choice of our test galaxies was to host
faint AGNs.
Specifically, their central sources have a reported luminosity of
log$(L_{X} [erg~s^{-1}]) = 39.2$ (NGC~4552) and log$(L_{X} [erg~s^{-1}]) = 38.8$
(NGC~4494) in the 2--10~keV band \citep{ogm:2009}, classifying them as LLAGNs.
Moreover, for NGC~4552 \cite{xu:2005} report an X-ray variability timescale of
$\sim$1~hr, while  \cite{maoz} and \cite{cappellari:1999} report long-term
variability in the UV bands.
Finally, in their analysis of the 1D surface brightness profile of NGC~4552
in the \HST{} bands, \cite{cappellari:1999} and
\cite{dullo:2014} fit the central bump with a point-like component
(although in Section \ref{Discussion} we argue this is an
artefact of dust extinction).

The images for the sample galaxies were retrieved from the Hubble
Legacy Archive (HLA)\footnote{
 \url{https://hla.stsci.edu}
}: their sublime spatial resolution ($\sim$0.15$\arcsec$ Gaussian FWHM)
is ideal for the study of the small (tens of pc) dust structures we intended to
characterize.
We chose blue bands, in which the effects of dust extinction are maximized;
nominally the ACS/$F475W$ and WFPC2/$F555W$ filters for NGC~4552 and NGC~4494,
respectively.
We sought for the image with the largest exposure time in the
respective band, however the deepest image for NGC~4494 in the $F555W$
filter is corrupted right at the center of the dust ring (probably due
to an incorrect exposure combination), hence we used the second best.
The details of the data set are provided in Table \ref{table:sample_images}.

\section{Modelling of dust rings}
\label{Modelling of dust rings}

\noindent
Our technique for the modelling of dusty structures\footnote{
 In the reminder we will use the term `ring' to indistinctly refer
 either to a ring, a disk, or a toroidal dust structure, unless the actual geometry
 is specified.
} is composed of two steps.
The first requires to parametrize the `pure' stellar profile of the galaxy, i.e.\
as if the dust component was not present (Section \ref{Modelling of galaxian light profile}).
In the second step, different dust geometries and masses are embedded within
the pure stellar profiles, generating a library of dust-absorbed models.
These models, generated through radiative transfer prescriptions, are then fitted
(scaled in intensity and matched in x,y position) to the original galaxy image.
The best fitting model automatically selects the bona-fide dust properties
(Section \ref{Monte-Carlo simulation of dust geometry and 2D fitting}).
The stellar surface brightness fitting and the dust-absorbed model fitting are both
performed in two dimensions (2D), in order to allow us to account for the rotation
angles of the dust rings.
The obvious name of choice for our pipeline --- in line with the geeky formalism of
astronomers --- was `Lord Of The Rings' (LOTR).

\subsection{Modelling of galaxian light profile}
\label{Modelling of galaxian light profile}

\begin{table*}
 \centering

 \begin{tabular*}{0.8\textwidth}{l @{\hskip 0.3cm} c @{\hskip -0.1cm} ccccccc}
  \hline
  \multicolumn{9}{c}{\textsc{Parameters of the stellar distributions}} \\
  \hline
  \hline
  \addlinespace 
  \multicolumn{1}{c}{Target}     & Best-fit Model & & Comp.       & S\'{e}rsic & $R_{e}$    & $R_{e}$    & Comp. $m$  & Comp. L\\
                                 &                & & Model       & $n$        & [arcsec]   & [kpc]      & [ABMAG]    & [L$_{\odot}$]\\
  \multicolumn{1}{c}{\tiny{(1)}} & \tiny{(2)}     & & \tiny{(3)}  & \tiny{(4)} & \tiny{(5)} & \tiny{(6)} & \tiny{(7)} & \tiny{(8)} \\
  \hline
  \addlinespace 
NGC~4552                  & \Sersic{}                       &                           & \Sersic{}   & 4.3 & 21.79 & 2.27 & 10.70 & $1.32 \times 10^{10}$\\
  \addlinespace 
\multirow{2}{*}{NGC~4494} & \multirow{2}{*}{\Sersic{}+exp.} & \multirow{2}{*}{\LARGE\{} & \Sersic{}   & 2.3 & 3.63  & 0.23 & 12.36 & $1.69 \times 10^{9}$\\
                          &                                 &                           & exponential & 1*  & 24.25 & 1.56 & 10.84 & $6.91 \times 10^{9}$\\
  \hline

 \end{tabular*}
 \parbox{0.8\textwidth}{
  \caption[]{
   \GALFIT{} best-fit parameters obtained from the fit to the pure stellar
   light distributions (i.e.\ excluding the areas affected by the absorption
   by the central dust rings).
   \\
   (1) Target name.
   (2) Best-fit total model.
   (3) Component model.
   (4) \Sersic{} index; (*) the exponential profile is equivalent to a \Sersic{}
       function with a fixed index $n = 1$.
   (5) Effective radius in units of arcseconds.
   (6) Effective radius in units of kilo-parsecs.
   (7) Component magnitude.
   (8) Component luminosity in solar units; the solar luminosity is expressed
            here following the formalism of \SKIRT{} (see Section
            \ref{Monte-Carlo simulation of dust geometry and 2D fitting}), i.e.\ as
            the solar luminosity at the pivot wavelength of the filter
            (nominally:
             L$_{\odot}$ = $2.78 \times 10^{33}$ erg/sec @ 0.475~$\mu$m, and
             L$_{\odot}$ = $2.94 \times 10^{33}$ erg/sec @ 0.555~$\mu$m).
   \label{table:GALFIT} 
  }
 }
\end{table*}

\noindent
The software of choice for the 2D fitting of the surface brightness of the sample
galaxies was \GALFIT{} \citep{GALFIT}.
An important caveat is that, in order to infer the pure stellar profiles, the
central areas affected by dust absorption are to be accurately masked.
Differently from the `canonical' approach of solely masking the dust features
(i.e.\ the ring) and leaving the innermost bright areas un-masked, we rejected the
whole area contained within the outermost edge of the ring.
This is because the luminosity of the bright areas inside the ring does not
only depend on the stellar/AGN light, but it is also determined
by the geometry of the dust (e.g.\ the size of the inner hole) and by the
reflection on the ring itself.
It is therefore not correct to use central data to fit parametric models accounting
solely for stellar/AGN emission.
Unfortunately, many past studies inaccurately overlooked this caveat, with the
result of incorrectly identifying non-existing galaxy cores or central point sources,
as discussed in Section \ref{Discussion}.

Apart from the complete masking of the central dust feature, the rest of the
\GALFIT{} analysis reflected the `standard' 2D fitting routine
\citep[see e.g.][]{bonfini:A2261}.
In brief, necessary corollary images are: the point spread function (PSF), the
uncertainty (`sigma') map, and the contaminant objects mask.
For the PSFs, we used the TinyTim tool \citep{TinyTim}, which allows to generate
ray-traced point spread functions for the \HST{} instruments; in particular,
we generated the PSFs corresponding to the pixel positions corresponding
to the center of the sample galaxies, to which we were most interested.
In our experience, the TinyTim PSFs usually slightly underestimates the seeing
by a few percent, but in our images there were not enough suitable stars to produce
a data-generated PSF.
The sigma images were generated by the internal Poissonian algorithm of \GALFIT{}, to
which we provided the background level and root-mean-square (RMS), which we measured on
background images created through \textsc{SExtractor} \citep[e.g.][]{sextractor}.
Contaminant objects were identified and masked starting from the \textsc{SExtractor}
`segmentation' maps: these are FITS images in which every pixel deemed part of
a source is masked out, essentially creating a map of the detected objects.
We run \textsc{SExtractor} twice (and combined the results), first tuned to
identify extended contaminant objects, and then to detect smaller objects embedded in
the galaxy light.
Through a \textsc{Perl} script\footnote{
 Publicly accessible from:
 \url{https://paolobonfini.wordpress.com/2016/05/04/mask-borders-of-a-fits-image}
}
we expanded isometrically the borders of each object in the segmentation images,
hence aggressively masking all possible contaminants.
Such masks were integrated with the dust masks described above, and with the
hand-made masks for problematic objects.

The fitting box was fixed to 400$\arcsec\times$400$\arcsec$ for both galaxies.
This range was chosen as the best compromise between the ability of correctly
reconstructing the galaxian light profile (whose inward extrapolation will,
in the next step, be affected by dust absorption), and the computational time
required by the radiative transfer code to reproduce images of the same size
(see Section \ref{Monte-Carlo simulation of dust geometry and 2D fitting}).
In this context, it is important to remember that the single orbit HLA mosaics of
nearby galaxies are known to be background over-subtracted, due to the relative
extent of the galaxies with respect to the field of view of the cameras.
Additionally, the WFPC2 mosaics suffer of the additional issue of
bias discontinuity between its 4 CCDs \citep[see discussion in][]{bonfini:corsair}.
Both of these features manifest as sharp turn-downs in the light profiles at large
radii, which we avoid with our choice of fitting box.

For NGC~4552 we fit a simple \Sersic{} component (under the assumption that the
galaxy does not host a core nor a central AGN), while for NGC~4494 we adopt a
\Sersic{} + exponential halo component.
The choice for a single/double component model is strongly supported
by the kinematics of the galaxies: while NGC~4552 shows a homogeneous
velocity dispersion pattern \citep{krajnovic:2011}, NGC~4494 present a double
structure \citep{foster:NGC4494}.
The best-fit parameters are reported in Table \ref{table:GALFIT}.

\subsection{Monte-Carlo simulation of dust geometry and 2D fitting}
\label{Monte-Carlo simulation of dust geometry and 2D fitting}

\begin{table*}
 \centering

 \begin{tabular*}{0.95\textwidth}{l @{\hskip 0.3cm} ccccccccccccc}
  \hline
  \multicolumn{14}{c}{\textsc{Parameters of the dust distributions}} \\
  \hline
  \addlinespace 
  \hline
  \addlinespace 
   \multicolumn{1}{c}{Target} &
   Dust                       &
   FOV                        &
                              &
                              &
                              &
   \multicolumn{3}{c}{\hrulefill~~\emph{disk}~~\hrulefill}  &
   \multicolumn{4}{c}{\hrulefill~~\emph{torus}~~\hrulefill} &
                              \\
                              &   
   model                      &
   P.A.*                      &   
   $\alpha$*                  &
   $\beta$*                   &
   $\gamma$*                  &
   $h_r$                      &
   $h_z$                      &
   $r_{i,d}$                  &
   $p$*                       &
   $r_{i,t}$                  &
   $r_{o,t}$                  &
   $\Delta\theta$             &
   $M_{dust}$                 \\
   \multicolumn{1}{c}{} &
                        &
   [deg]                &
   [deg]                &
   [deg]                &
   [deg]                &
   [pc]                 &
   [pc]                 &
   [pc]                 &
                        &
   [pc]                 &
   [pc]                 &
   [deg]                &
   [log($M/M_{\odot}$)] \\
   \multicolumn{1}{c}{\tiny{(1)}} &
   \tiny{(2)}                     &
   \tiny{(3)}                     &
   \tiny{(4)}                     &
   \tiny{(5)}                     &
   \tiny{(6)}                     &
   \tiny{(7)}                     &
   \tiny{(8)}                     &
   \tiny{(9)}                     &
   \tiny{(10)}                    &
   \tiny{(11)}                    &
   \tiny{(12)}                    &
   \tiny{(13)}                    &
   \tiny{(14)}                    \\
  \hline
  \addlinespace 
NGC~4552 & disk  & 0  & 40 & 154 & 25 & 5.2$^{{\tiny \ ^{\vee}}}_{-2.4}$ & 0.5$^{+0.8}_{{\tiny \ _{\wedge}}}$ & 0.3$^{+0.2}_{-0.2}$ & -- & -- & -- & -- & 2.9$^{+0.8}_{{\tiny \ _{\wedge}}}$ \\
  \addlinespace 
NGC~4494 & torus & 12 & 0  & 120 &  0 & --  & --  & --  & 1  & 32$^{+4}_{-12}$ & 41 $^{+16}_{{\tiny \ _{\wedge}}}$ & 9$^{+4}_{-4}$ & 2.9$^{+0.3}_{-0.9}$ \\
  \addlinespace 
  \hline

 \end{tabular*}
 \parbox{0.95\textwidth}{
  \caption[]{
   Best-fit parameters obtained from the fit of the dust-absorbed model libraries
   generated by \SKIRT{} to a galaxy image.
   Parameters held `frozen' in our procedure are marked with a star (*).
   Upper and lower limits are marked with the symbol $\wedge$ and $\vee$, respectively.
   \\
   (1)  Target name.
   (2)  Dust model geometry; notice that the term `torus' refers to the geometry
        chosen among the \SKIRT{} models, not to the torus surrounding the AGN
        (see Equation \ref{equation:sublimation} and relevant discussion).
   (3)  Position angle of the instrument simulated by \SKIRT{}; this can be used to align
        the P.A. of the dust feature with the y-axis of the image (and hence allow the usage
        of one less Euler angle).
   (4,5,6) \SKIRT{} Euler angles defining the orientation of the dust geometry;
        please refer to the user's manual for the non-canonical definition of
        Euler angles in \SKIRT{}.
   (7)  Disk radial scale length.
   (8)  Disk vertical scale height.
   (9)  Disk inner truncation (hole) radius.
   (10) Torus radial density power-law index.
   (11) Torus inner radius.
   (12) Torus outer radius.
   (13) Torus opening angle.
   (14) Dust mass.
   \label{table:fit} 
  }
 }
\end{table*}

\noindent
The 2D stellar parametric models obtained as described in Section \ref{Modelling of galaxian light profile}
were used as input for the radiative transfer software, through which we added the
dust component and re-calculated the expected light emission.
To this purpose, we used \SKIRT{} \citep[e.g.][]{SKIRT}, a software which emulates
the physical processes involving
dust (scattering, absorption and emission) through a Monte Carlo technique.
\SKIRT{} allows the user to set-up two types of classes, i.e.\ the stellar and
the dust components.
Furthermore, the geometry of both classes can be modified in several ways by
applying a set of ``decorators''.
A decorator is a way that \SKIRT{} uses to implement changes in a given geometry
(e.g.\ a ``spiral arm'' decorator applied to a disk geometry, will modified the
density distribution of a disk in such a way to mimic logarithmic spiral arms.
Rotating a given geometry is another example of a decorator)\footnote{
 The description of all available classes/decorators is accessible online, on
 the project website: \url{www.skirt.ugent.be/root/index.html}.
}.

In developing our pipeline we first attempted to adopt the publicly available Fit\SKIRT{}
\citep{FitSKIRT};
this is a genetic algorithm-based wrapper for \SKIRT{} designed to automatically explore
the parameter space and find the dust best-fitting structural parameters.
However in our case, due to a very high degeneracy between the disk scale-heights and
the inclination angles, using Fit\SKIRT{} turned out to be infeasible.
In fact, either the fits were not converging, or the parameters for the dust disks
turned out to be unphysical.
The reason for this poor performance arguably lies in the fact that Fit\SKIRT{} has
been constructed, tested, and optimized specifically to study edge-on spiral galaxies.
We therefore designed our own approach, as described in the following.

To simulate our \Sersic{} (+ exponential halo) stellar components we used
the \SKIRT{} built-in \Sersic{} class\footnote{
 Note that an exponential function is equivalently described by a \Sersic{} with
 index $n$ = 1.
}, to which we applied a `rotation' and a `triaxial'
decorator to reproduce the observed position angle (P.A.) and ellipticity ($e$),
respectively.
In this context, we were only interested in reproducing the projected appearance of the
galaxy rather than its three-dimensional geometry, to which the \SKIRT{}
decorators formally applies.
At the center of these stellar components we added the \SKIRT{} dust components.
Their geometries were chosen after a careful visual inspection of the dust
features and of their reflection patterns, and after testing with different
models.
For NGC~4552 we adopted the \SKIRT{} `disk' component, of the form:

\vspace{-0.5cm}
\begin{eqnarray}
 \rho(r,z) = \rho_{0} ~ \exp\Bigl\{-(r/h_r + z/h_z)\Bigr\} & for & r > r_{i,d}
 \label{equation:disk}
\end{eqnarray}

\noindent
where $h_{r}$ and $h_{z}$ are the radial and vertical scale lengths, $r_{i,d}$ is
the size of the inner truncation, and  $\rho_{0}$ is the central density
(equivalently determined by the total dust mass $M_{dust}$), i.e.\ the innermost
density of the exponential profile was it not truncated by the inner hole.
For the case of NGC~4494 we opted instead for the \SKIRT{} `torus'\footnote{
 In this context, the term `torus' does not refer to the AGN torus, but rather to
 the \SKIRT{} model class used to reproduce a generic dust ring.
}:

\vspace{-0.5cm}
\begin{eqnarray}
 \rho(r,z) = A ~ r^{-p} e^{-q|cos\theta|} & for & r_{o,t} > r > r_{i,t} \\
                                          & and & {\pi \over 2} - \Delta\theta < \theta < {\pi \over 2} + \Delta\theta\nonumber
 \label{equation:torus}
\end{eqnarray}

\noindent
characterized by the inner ($r_{i,t}$) and outer ($r_{o,t}$) radii, the radial density
power-law index $p$, the polar index $q$, the opening angle $\Delta\theta$ (i.e.\ the
vertical angular extent as seen from the center of the torus), and the normalization
parameter $A$ (equivalently determined by $M_{dust}$).
Dependencies of the dust density on the polar angle where ignored by keeping the $q$
parameter fixed to 0.
For both galaxies we adopted a dust mixture following the prescriptions of
\cite{zubko}: these account for a composition of graphite, silicate, and neutral and
ionized PAH dust grains, and are calibrated over the dust emission, extinction and
abundance of dust in the Milky Way. 

\begin{figure*}
 \centering

 \begin{minipage}[b]{0.32\textwidth}
  \centering
  \textsc{\large Data}
 \end{minipage}
 \begin{minipage}[b]{0.32\textwidth}
  \centering
  \textsc{\large Model}
 \end{minipage}
 \begin{minipage}[b]{0.32\textwidth}
  \centering
  \textsc{\large Residuals}
 \end{minipage}
  
 \vspace{0.2cm}


 \makebox[\linewidth]{
  \includegraphics[align=c,width=0.3\textwidth]{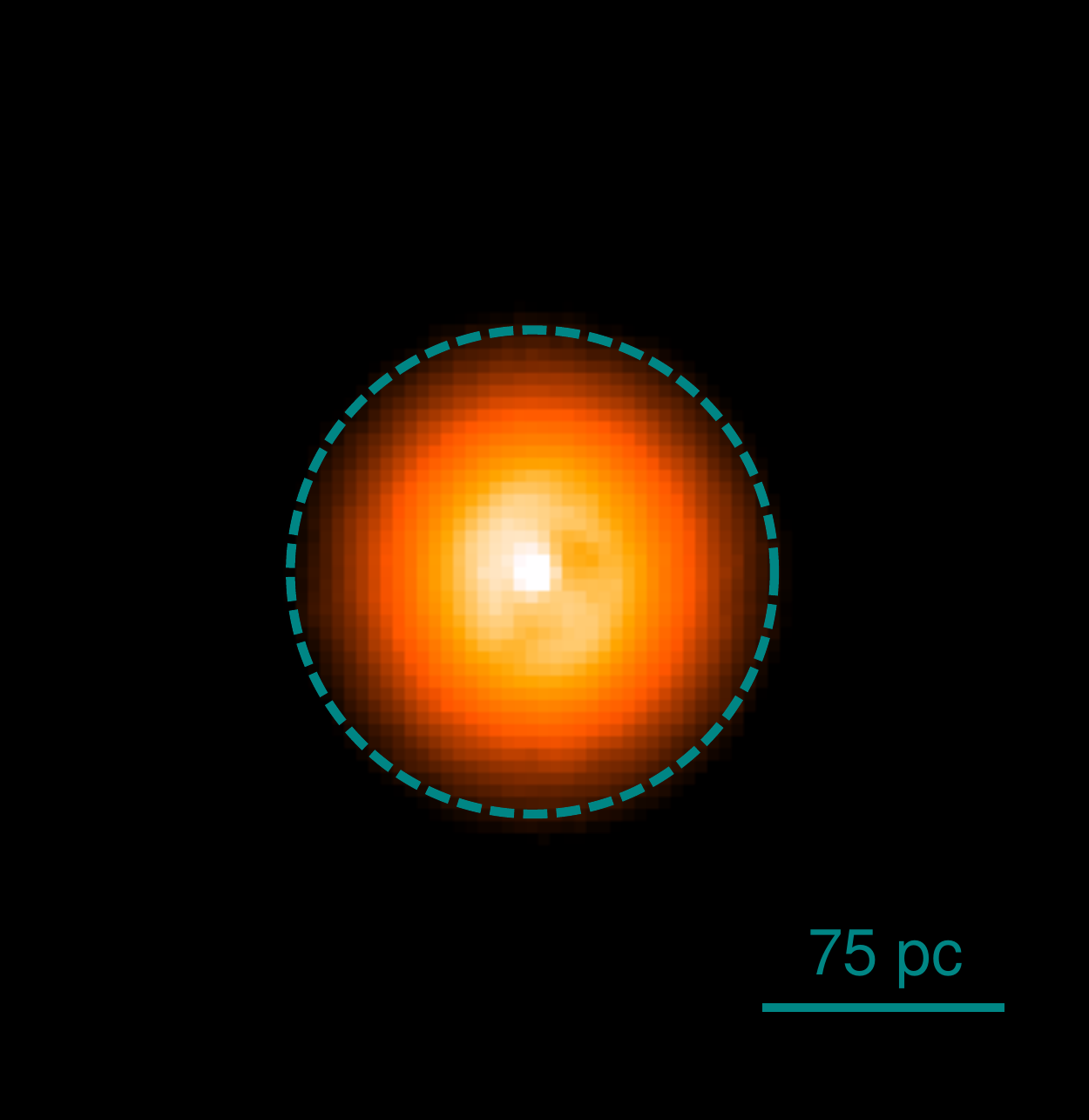}
  
  \begin{minipage}[c]{0.3\textwidth} 
   \begin{overpic}[align=c,width=1\textwidth]
    {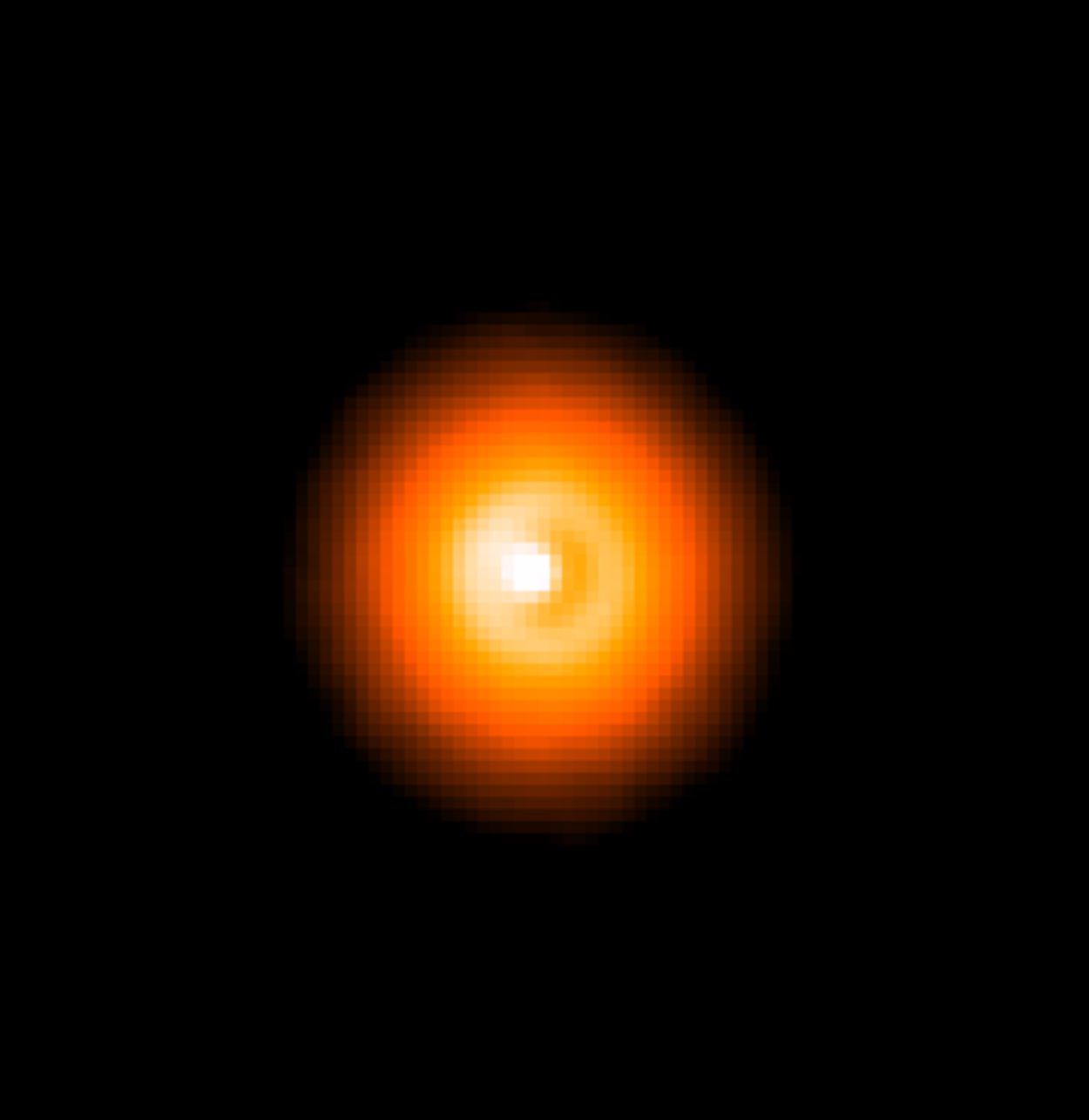}
    \put(50,10){\makebox(0,0){\textcolor{white}{\large NGC~4552}}}
   \end{overpic}
  \end{minipage}
  
  \begin{minipage}[c]{0.3\textwidth} 
   \begin{overpic}[align=c,width=1.3\textwidth, height=1.03\textwidth]
    {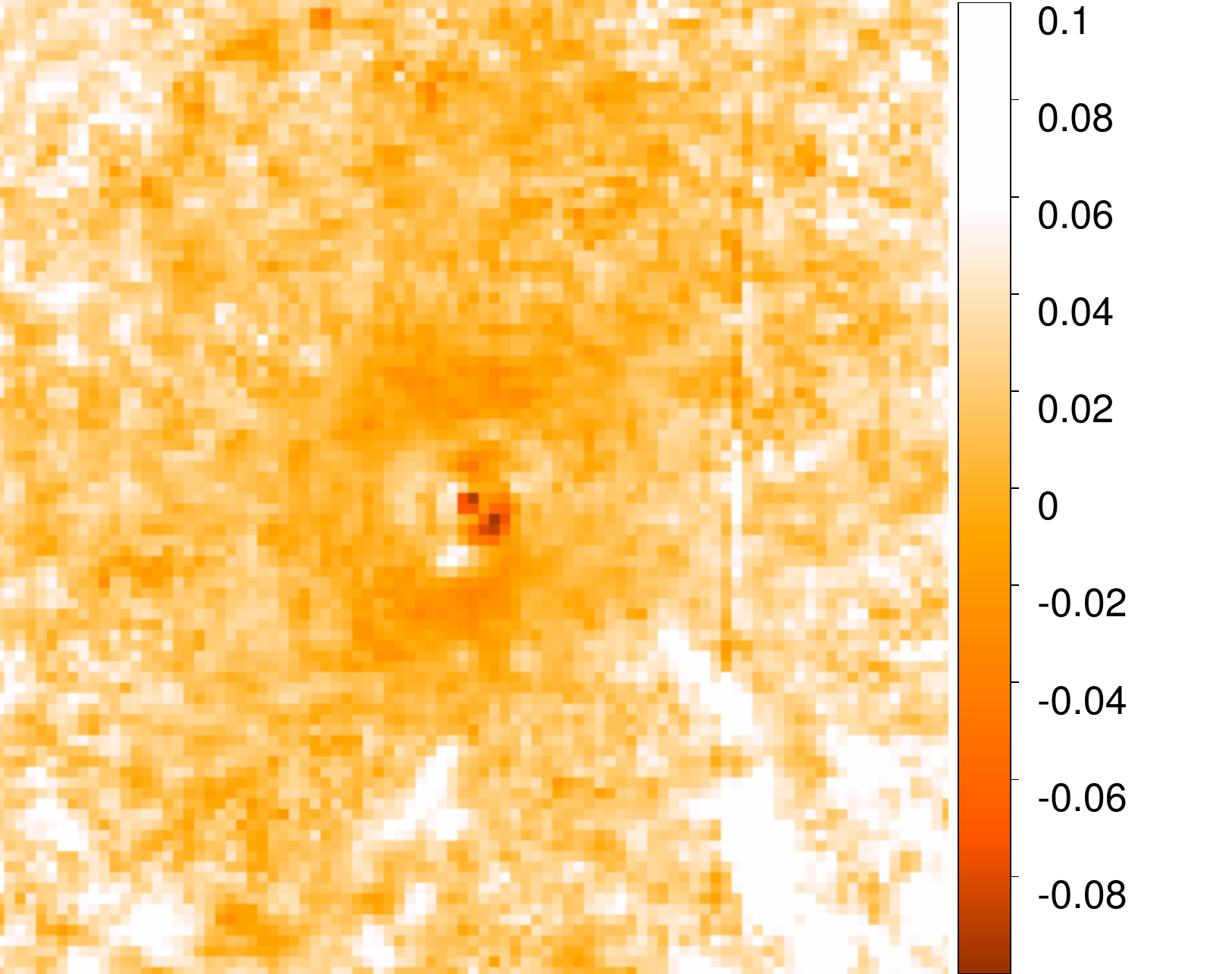}
    \put(95,40){\makebox(0,0){\rotatebox{-90}{\textcolor{black}{\large $\Delta\mu_{F475W}$}}}}
   \end{overpic}
  \end{minipage}

  
 }
 
 \vspace{0.2cm}

 \makebox[\linewidth]{
  \includegraphics[align=c,width=0.3\textwidth]{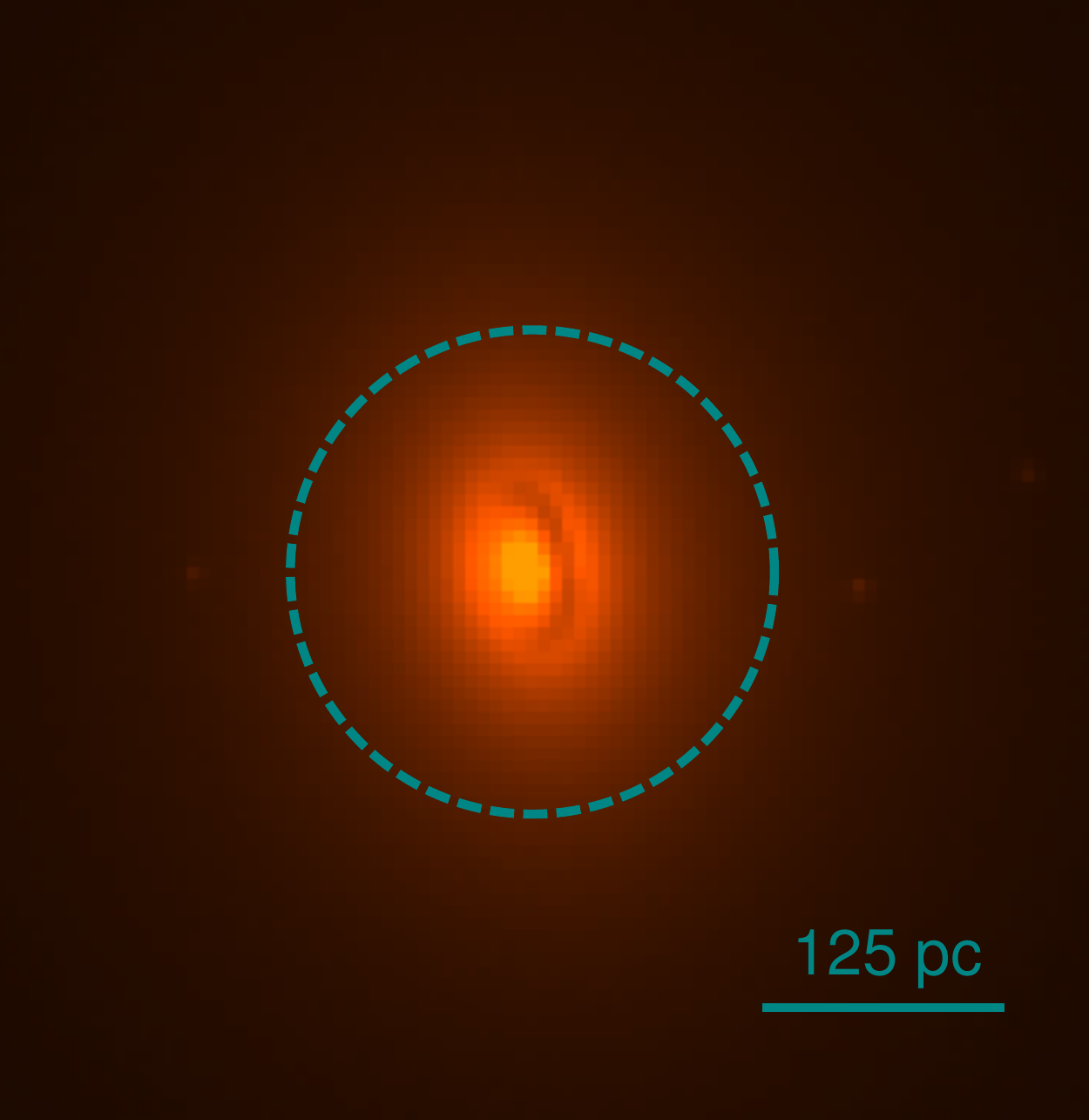}
  
  \begin{minipage}[c]{0.3\textwidth} 
   \begin{overpic}[align=c,width=1\textwidth]
    {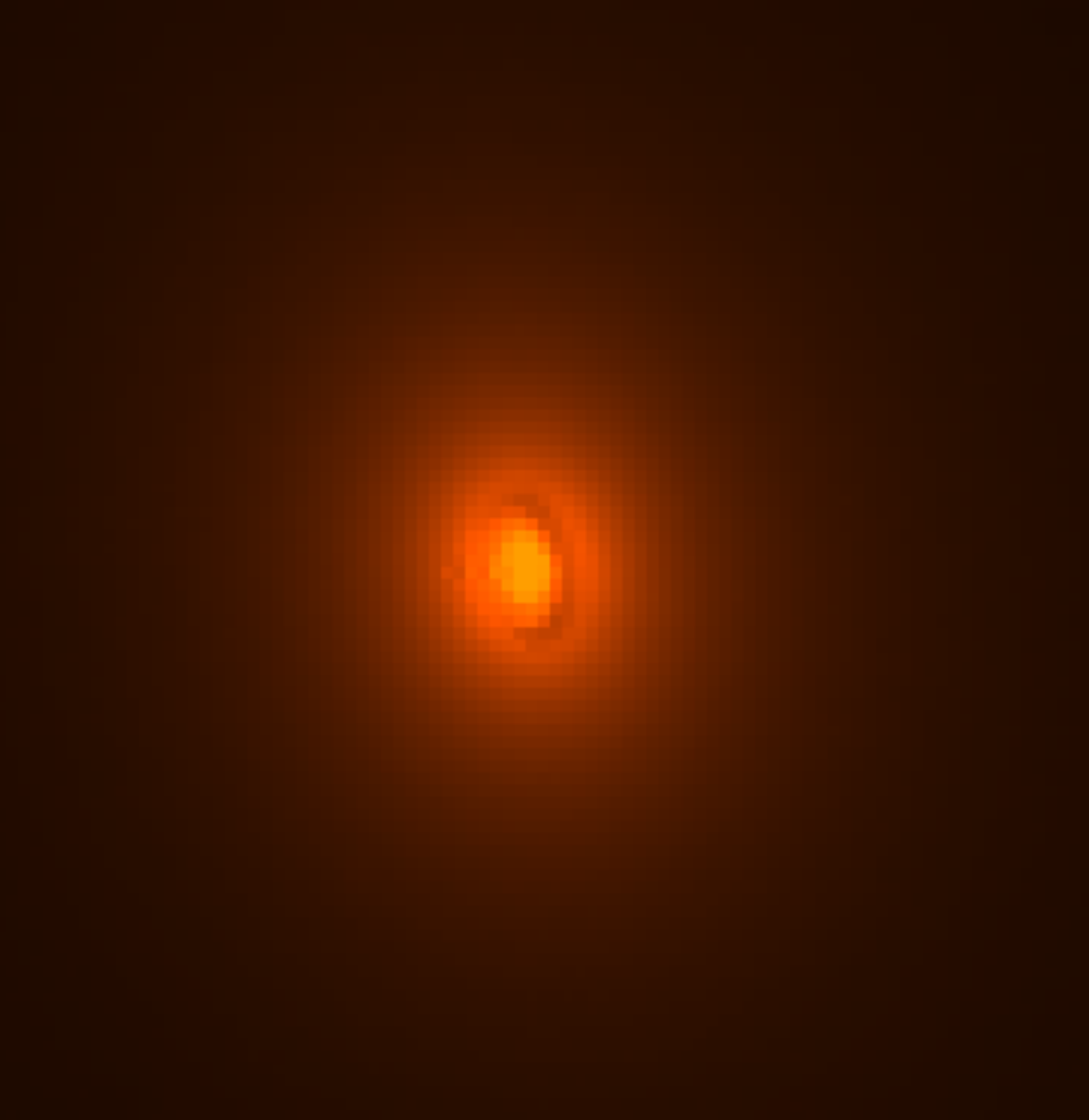}
    \put(50,10){\makebox(0,0){\textcolor{white}{\large NGC~4944}}}
   \end{overpic}
  \end{minipage}
  
  \begin{minipage}[c]{0.3\textwidth} 
   \begin{overpic}[align=c,width=1.3\textwidth, height=1.03\textwidth]
    {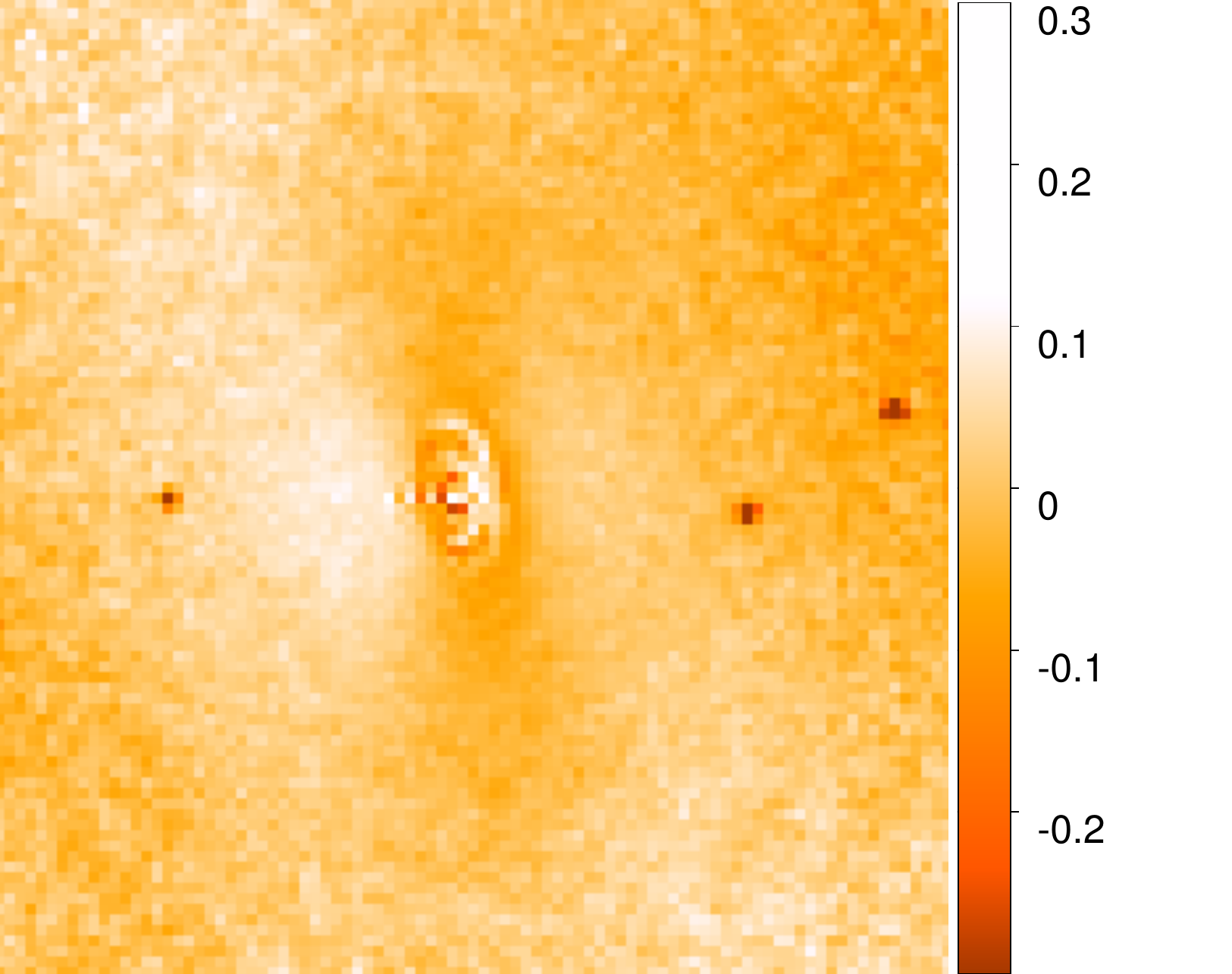}
    \put(95,40){\makebox(0,0){\rotatebox{-90}{\textcolor{black}{\large $\Delta\mu_{F555W}$}}}}
   \end{overpic}
  \end{minipage}
  
  
 }
 
 \vspace{0.2cm}
 
 \caption[]{
  Final results of our LOTR fitting pipeline for NGC~4552 (\emph{top}), and NGC~4494
  (\emph{bottom}).
  \emph{Left ---} Original HLA images.
                  The dashed circles represent the areas over which the fitting
                  of the dust-absorbed models was performed, and correspond to
                  an angular radius of  1$\arcsec$ for NGC~4552, and 2$\arcsec$
                  for NGC~4494.
                  They also roughly correspond to the areas masked in the first step
                  of LOTR (Section \ref{Modelling of galaxian light profile})
                  during which we were interested in modelling the stellar emission
                  avoiding all possible extinction by dust.
  \emph{Center ---} Best-fit dust-absorbed model generated through \SKIRT{} using
                  the stellar components determined from the surface brightness
                  fit to the galaxies (Table \ref{table:GALFIT}; Section
                  \ref{Modelling of galaxian light profile}), and the
                  best-fit dust components from our Monte-Carlo simulation
                  (Table \ref{table:fit}; Section
                  \ref{Monte-Carlo simulation of dust geometry and 2D fitting}).
  \emph{Right ---} Data minus model residuals expressed in terms of surface
                  magnitude difference ($\Delta\mu$).
                  The color map scaling has been chosen in order to match
                  the corresponding $\Delta\mu$ range in the bottom panel of Figure
                  \ref{figure:projections}, from which it is clear that
                  the represented surface brightness fluctuations are well below
                  0.1~mag/arcsec$^2$.
                  The bright ``stripes'' in the residual image for NGC~4552
                  are not artefacts, but rather correspond to real dust lanes.
  \\
  \textsc{Note:} The images for NGC~4494 refer to the model including the AGN
                 component (blue data series in Figure \ref{figure:projections}).
 }
 \label{figure:fit_models}

\end{figure*}

We visually chose and fixed the angles by which the central dusty structure is
rotated with respect to the line of sight of the observer.
In principle, rotation angles are degenerate with
the disk/torus thickness, but since the structures turned out to be physically
thin (see Table \ref{table:fit}), the error we introduced by locking the angles
was marginal and was abundantly compensated by a significant speeding-up of the
simulations.
All the other dust parameters were allowed to vary with the exception of the
torus index $p$, which we set to 1.
We tentatively chose first guess parameters by visually inspecting the images
and producing a number of models to get fairly close to the observed surface
brightness profile.
Hence, we defined a grid in the parameter space that we sampled with discrete
values by varying each parameter to within a factor of 2 from
its initial best guess.
For each combination of parameters we created a \SKIRT{} model,
which we subsequently convolved with the TinyTim PSF (Section \ref{Modelling of galaxian light profile})
corresponding to the seeing of the instrument we intended to simulate,
hence creating a `library' of dust-absorbed models.
Each sampled model was then fit to the original galaxy image using \GALFIT{}, to
which we provided --- for consistency --- the same error images utilized during the
previous surface brightness fitting step (Section \ref{Modelling of galaxian light profile}).
However, for this part of the procedure we used a mask which left uncovered only an
area centered on the ring and roughly doubling its size (dashed circles in
Figure \ref{figure:fit_models}), since we wanted to focus our fit on the central region.
In fact, since the parametric stellar component provided to \SKIRT{} is exactly what
was fitted in the real image, it automatically reproduced the outer galaxian light.
To obtain the best-fit model, we simply performed a \Chisq{} search of the
sampled parameter space.

The parameter uncertainties were estimated based on the RMS of the residuals,
according to the following procedure.
First we evaluated --- for the best-fit model --- the `on-source' RMS ($RMS^{best}_{S}$),
i.e.\ the RMS of the residuals within the fit area, and the `background' RMS
($RMS^{best}_{B}$), i.e\ the RMS evaluated over the pixels outside the fit area.
Then, we selected a pool of models whose on-source RMS ($RMS_{S}$) was such that:

\vspace{-0.5cm}
\begin{eqnarray}
 RMS_{S} < RMS^{best}_{S} + 3 \times RMS^{best}_{B}
 \label{equation:RMS}
\end{eqnarray}

\noindent
i.e.\ less than three times above the residual background noise.
Finally, the upper and lower uncertainties for each parameter were respectively
defined as the maximum and minimum values of that parameter within the
pool of selected models.
Due to the relatively scarce sampling of the parameter space, dictated by the required
computational time,
for several parameters we could only identify upper or lower limits.
The best-fit parameters for the dust components, and their uncertainties, are
reported in Table \ref{table:fit}.

\begin{figure*}
 \centering

  \makebox[\linewidth]{

  \begin{overpic}[align=c,width=0.48\textwidth]
   {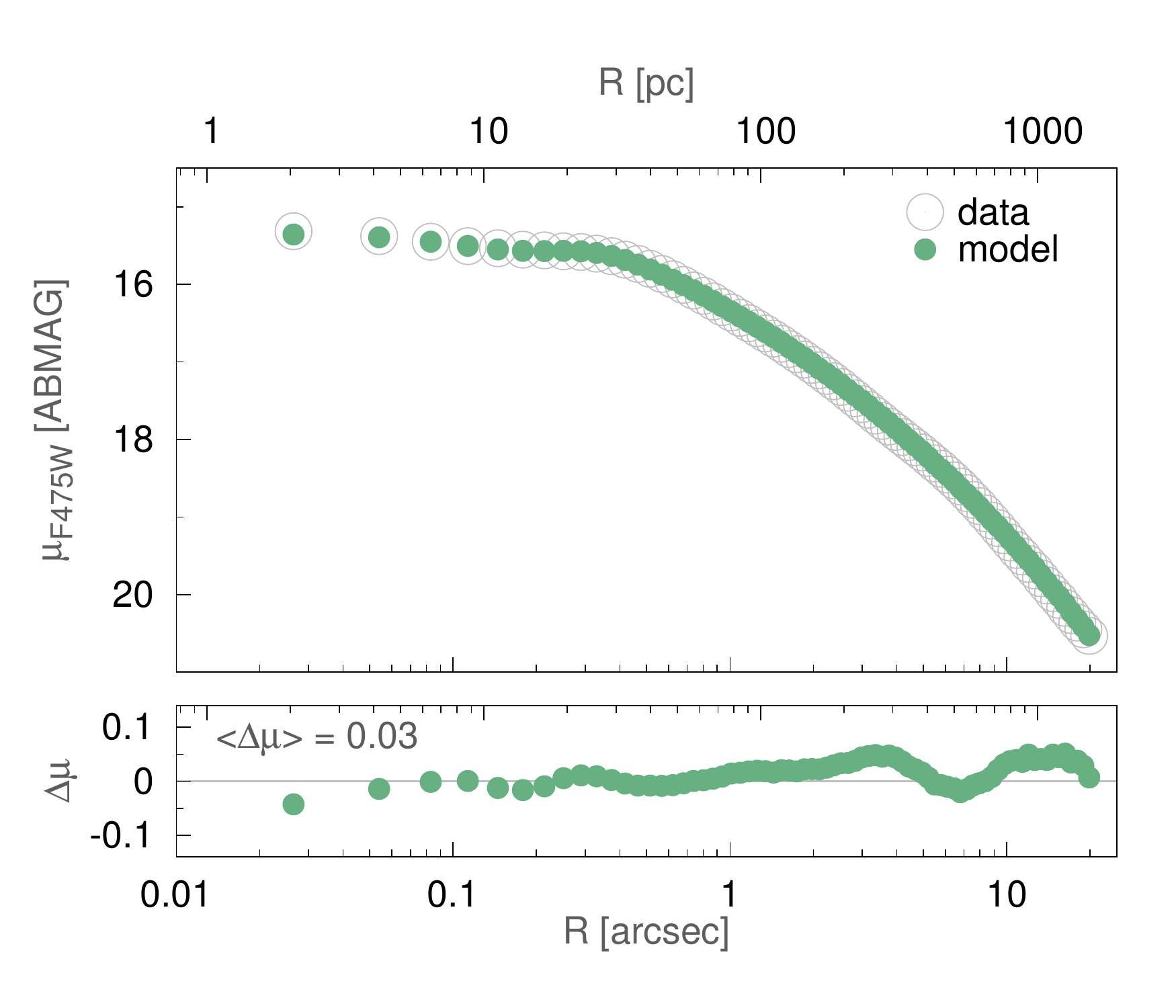}
    \put(30,35){\makebox(0,0){\textcolor{black}{\large NGC~4552}}}
   \end{overpic}

  \begin{overpic}[align=c,width=0.48\textwidth]
   {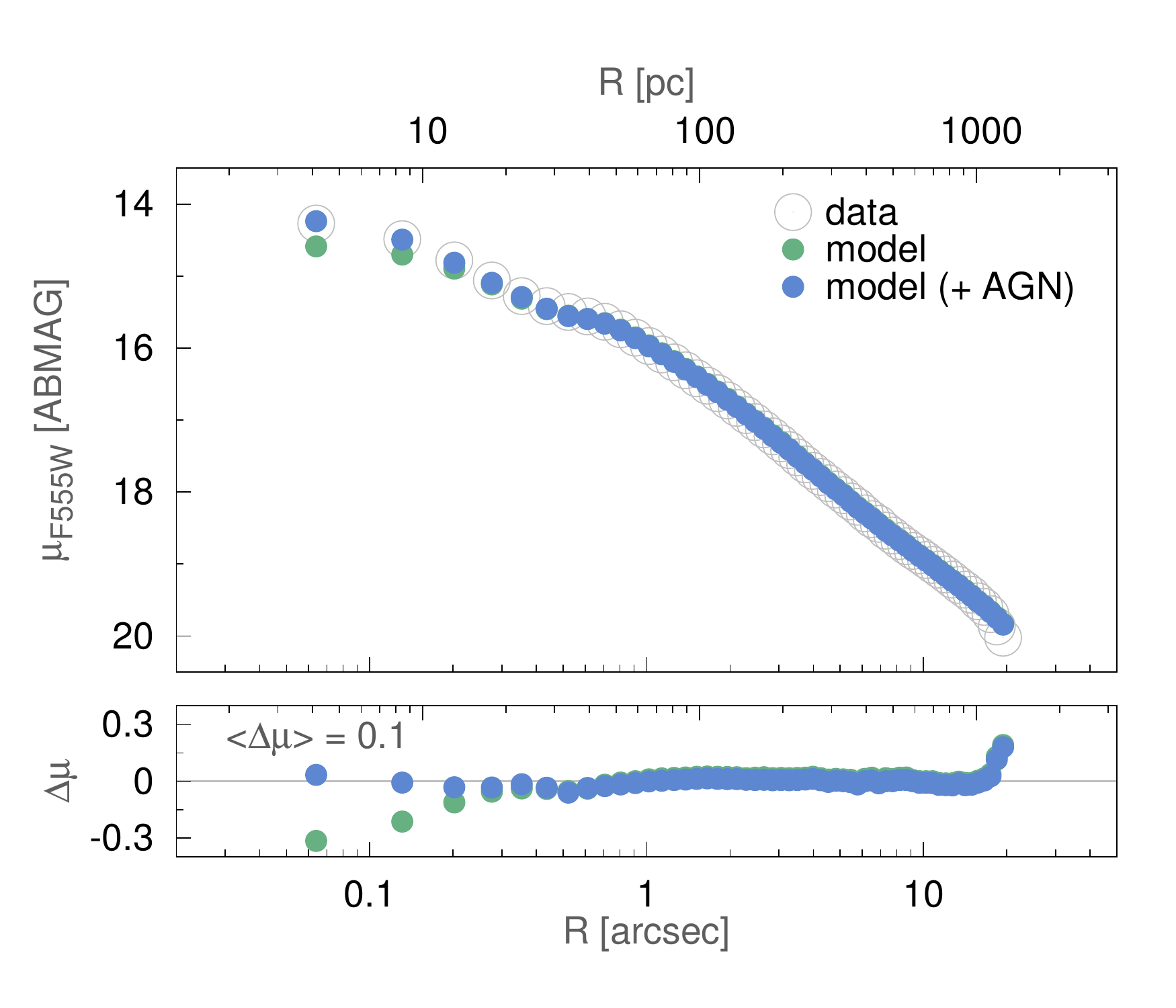}  
   \put(30,35){\makebox(0,0){\textcolor{black}{\large NGC~4944}}}
  \end{overpic}

 }
  
 \caption[]{
  Projection of the 2D data, model, and residual images of NGC~4552 (\emph{left}),
  and NGC~4494 (\emph{right}), performed over concentric circular annulii.
  We stress that these plots do not represent the result of a fit to the 1D data series,
  but rather an azimuthally  averaged representation of Figure \ref{figure:fit_models}.
  The projections extend further than the images of Figure \ref{figure:fit_models},
  and they actually cover the whole radial range fitted with \GALFIT{} in the first step of
  the LOTR pipeline (Section \ref{Modelling of galaxian light profile}).
  \emph{Top ---} Surface brightness profiles of the data (empty circles)
                 and of the best-fit models identified with our pipeline
                 (green circles).
                 For NGC~4494, we additionally show (blue circles) the profile
                 generated by adding an arbitrarily luminous AGN at the center
                 of the best-fit dust-absorbed model, and re-performing the 2D
                 fit.
  \emph{Bottom ---} Surface magnitude difference ($\Delta\mu$).
                 The $\Delta\mu$ RMS value $\left\langle\Delta\mu\right\rangle$
                 is indicated in the plot, and in the case of NGC~4494 it refers
                 to the model without the AGN component. 
 }
 \label{figure:projections}

\end{figure*} 

In Figure \ref{figure:fit_models} we show the original, the best-fit model, and
the residual (data $-$ model) images.
A 1D projection of the results is presented instead in Figure \ref{figure:projections}.
For NGC~4494, the best-fit model selected by LOTR (including exclusively stellar
and dust components) underestimates the innermost ($R < 0.2\arcsec$) surface
brightness by $\sim$0.3~mag.
This discrepancy is most probably due to the presence of a faint central point source,
which we interpret as a LLAGN (Section \ref{Dust pretending to be a core or an AGN}).
To test this, we added a 5 $\times$ 10$^{6}$ L$_{\odot}$
($\sim$1.5 $\times$ 10$^{40}$~erg sec$^{-1}$) point source at the center of the 
dust-absorbed model, and then re-run the 2D fit in order to fine tune the new model
x,y position.
This modification yielded extremely flat residuals: in Figure \ref{figure:fit_models}
we show exactly this new model, while in Figure \ref{figure:projections} we compare
the models with and without AGN.
In principle, every emission component should be included at the iterative fitting
stage and its luminosity sampled along with the other parameters, in order to
account for any degeneracy with the dust properties.
However, since the inner radius of the torus is resolved
($r_{i,t}$ = 32~pc = 0.25$\arcsec$ $\sim$ 1.5 PSF FWHM), the AGN we added (point-like)
is not `shaded' by the dust.
Moreover, its modest luminosity does not produce any significant reflection
over the dusty torus.
Our analytical approach of treating the AGN as a post-sampling additional
component is therefore justified.

\section{Discussion}
\label{Discussion}

\subsection{Dust pretending to be a core or an AGN}
\label{Dust pretending to be a core or an AGN}

\begin{figure*}
 \centering

  \makebox[\linewidth]{

  \begin{overpic}[align=c,width=0.95\textwidth]
   {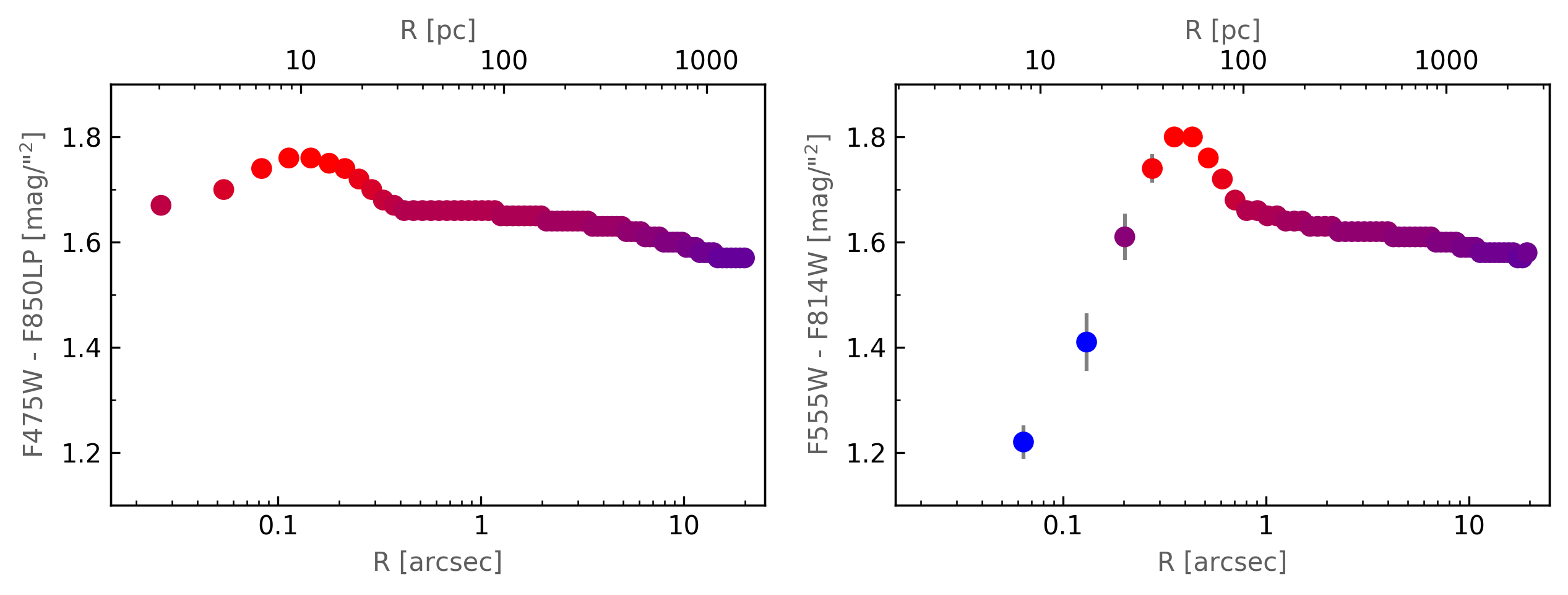}
    \put(40,10){\makebox(0,0){\textcolor{black}{\large NGC~4552}}}
    \put(90,10){\makebox(0,0){\textcolor{black}{\large NGC~4494}}}
   \end{overpic}

 }
  
 \caption[]{
  Radial colour profiles of the sample galaxies.
  The color coding is proportional to the colour value and meant to emphasize the
  reddening by the dust rings, which manifests as a red ``bump'' in the profile.
  The colours have been calculated in the original \HST{} filters, and they roughly
  correspond to the SDSS $g' - z'$ ($F475W - F850LP$) and Johnson-Cousins
  $V - I$ ($F555W - F814W$).
 }
 \label{figure:cmap}

\end{figure*} 

\noindent
The most surprising result of our analysis is the discovery that, once the effect
of dust is properly taken into account, NGC~4552 is \emph{not} compatible with
hosting a core, as previously claimed by \emph{several} studies.
Notably, the fact that NGC~4552 presents strong signs of `fine structure'
(i.e.\ indications of recent interactions \citealt{schweizer:1992}) is
potentially compatible with the lack of a [large] core \citep{bonfini:sigma_core}.
However, NGC~4552 has been classified as a `core' galaxy by \cite{faber:1997},
\cite{richings}, \cite{rusli}, \cite{dullo:2014}, apart
from the reference study we used to select our sample, i.e.\ \cite{lauer:2005}.
The core was confirmed even when the analysis included image correction
by dust extinction \citep[e.g.][]{ferrarese:2006}.
This indicates how easy it is to misclassify dust rings as cores --- but
how could this galaxy deceive so many investigations?
Excluding the early works, in which the concept of `core' was still under debate,
one possible explanation is that the aforementioned studies regarded large
samples, over which a detailed object-by-object investigation is not feasible.
However, we believe that the main source of discrepancy lays in their use of 1D
analysis.
With this technique, dust features are masked on the images and then the
surface brightness profiles extracted: since this procedure inevitably
averages the intensity over isophotes, it is admittedly difficult to keep
track of the (asymmetric) dust extinction at the fitting stage.

\cite{dullo:2014} measured a $V$-band point source magnitude $m_V = 20.6$~mag,
corresponding to $\sim$10$^{6}$~L$_{\odot}$, similar to what \cite{cappellari:1999}
measured in their $HST$-$F342W$ band.
From an other point of view, given that the point sources in NGC~4494 and NGC~4552
have similar X-ray emission (Section \ref{Sample and data}), and assuming a similar
X-ray to optical emission ratio, we can reasonably expect that they also
posses a similar optical luminosity, i.e.\ $\sim$5 $\times$ 10$^{6}$~$L_{\odot}$
(see Section \ref{Monte-Carlo simulation of dust geometry and 2D fitting}).
However, when we added a point source of such luminosity at the center of our
NGC~4552 best-fit model --- similarly to what we did for NGC~4494 ---
the fit residuals showed no statistical improvement.
We argue that the point source `detected' in the aforementioned studies was
an artefact of incorrectly assuming a core in NGC~4552.
In our physically simpler scenario, the central peak ($R < 0.15\arcsec$) is merely due to
stellar emission shining through the inner hole of the dust disk
(plus reflection of stellar light on the dust). 
In the UV band, in which the dust extinction is more significant, the contrast
between the radiation passing through the hole and that absorbed by the disk
is expected to be enhanced.
We hence believe that the fluxes of the UV `spikes' (as they named  the central
emission) measured by \cite{cappellari:1999} are particularly affected by this
problem.
In general, neglecting dust extinction can lead to a strong overestimation of the
luminosity of the AGN, or even to a false detection in optical/UV bands.

\cite{cappellari:1999} additionally studied the evolution of the central `spike'
of NGC~4552 during the period 1991--1996, and they recorded significant intensity
variation in the \HST{} UV bands
\citep[following on the UV flare observed by][]{renzini:NGC4552}, up to a factor
$\sim$4.5 (during the interval 1991--1993).
This figure is surprisingly high with respect to the fluctuations observed in
optical/UV surveys of AGNs \citep[e.g.][]{simm:2016}.
They suggest that this behaviour should be related to a sporadic accretion
event over the central black hole.
We argue that at least part of the variability that \cite{cappellari:1999}
measured was significantly affected by the temporal variations of the instrumentation
on board \HST{}.
Most notably, in between their 1991 and 1996 observations the Corrective Optics Space
Telescope Axial Replacement (COSTAR) was mounted and calibrated.
We wish to stress that --- although the authors took great care in calibrating their
frames --- these instrumental variations of the PSF and sensitivity could have alone
mimicked AGN variability.
Was the variability real, the suggested hypothesis of a sporadic accretion or even of a
UV flare from a star stripped from its envelope are not hypothesis at odds with our model:
the radiation released by these events can actually shine through the central hole of the dust ring,
similarly to the LLAGN in  NGC~4494; Section \ref{Monte-Carlo simulation of dust geometry and 2D fitting}).
However, these events are relatively rare occurrences. 
In our paradigm, variability could simply be due to a dust cloud crossing the line of
sight or to a change in the inner geometry of the dust ring, such as a variation in
`clumpiness', in the size of the hole, or in the dust density (although the timescales for
a change in the dust configuration would imply extremely high velocities).
For example, by (erroneously) fitting a core-\Sersic{} + PSF model to our best-fit
model, and to a model in which we just changed the best-fit vertical disk length
($h_{z} = 0.5$~pc) to its upper limit ($h_{z} = 1.3$~pc; Table \ref{table:fit}),
the flux encompassed by the PSF changes by a factor $\sim$2.5, close to the measurements
of \cite{cappellari:1999}.
This possibility is a simpler alternative to an AGN in explaining the optical/UV
variability observed in the central emission of NGC~4552.
In conclusion, despite NGC~4552 shows X-ray variability \citep{xu:2005} and emission
line indications for AGN activity \citep[e.g.][]{ho:1997,veron}, we argue
that this might be below detection levels in optical/UV imaging.

\medskip

\noindent
The case for NGC~4494 is less controversial.
Of the studies which attempted to fit both core and core-\emph{less} models to
their sample galaxies, only one \citep{rest} preferred not to doubtlessly
classify the galaxy as core-\emph{less}, due to the presence of the dust ring.
Our results further confirm that the central decrease in intensity
(with respect to the extrapolation of the outer light profile) is purely
due to extinction.
NGC~4494 hosts a LLAGN \citep[e.g.][]{veron} with log$(L_{X} [erg~s^{-1}]) = 38.8$
in the 2--10~keV band \citep[][]{ogm:2009}.
How does this luminosity compare to the $\sim$1.5 $\times$ 10$^{40}$~erg sec$^{-1}$
(or log$(L_{F555W}  [erg~s^{-1}]) = 40.2$) point source that we added to our
best-fit, dust-absorbed model in the $F555W$ filter (similar to the $V$-band)?
To perform a rough calculation, we can use the bolometric corrections derived by
\cite{elvis:1994} for bright AGNs: L$_{bol}$ $\sim$ 15 $\times$ L$_{X}$,
and L$_{bol}$ $\sim$ 13.2 $\times$ L$_{V}$, where L$_{bol}$ is the bolometric
luminosity, and L$_{X}$ and L$_{V}$ are the X-ray and $V$-band luminosities,
respectively.
Combined, these scaling relations yield: L$_{V}$ = 1.1 $\times$ L$_{X}$;
this is too little to explain the factor of 10 between our $V$-band luminosity
and the X-ray one.
However, the relations by \cite{elvis:1994} were calibrated over luminous
AGNs, and they might be very different for LLAGNs.
Unfortunately, even assuming that for LLAGNs L$_{bol}$ $\sim$ 30 $\times$ L$_{X}$
\citep[][their equation 21]{marconi:2004} and keeping valid the previous
L$_{bol}$--L$_{V}$ relation, we could reach L$_{V}$ $\sim$ 2 $\times$ L$_{X}$,
which is still quite far from our estimate.
One possible explanation compatible with our results is that the AGN is
Compton-thick: in that scenario one can expect up to
L$_{X,intrinsic}$ $\sim$ 60 $\times$ L$_{X,observed}$ \citep{panessa:2006}, which
would render the calculation broadly consistent with our L$_{F555W}$. 

\medskip

\noindent
As a further check for discerning the effects of dust extinction as opposed to
the presence of a core, we created color maps for the sample galaxies.
To do so, we additionally retrieved the $F850LP$  and $F814W$ images from the HLA
archive, and used them to produce maps in the $F475W - F850LP$ and $F555W - F814W$
colours, roughly equivalent to the SDSS $g' - z'$ and Johnson-Cousins $V - I$ colours,
respectively.
In Figure \ref{figure:cmap} we show a radial profile extracted from those colour
maps along the same concentric annulii utilized to produce Figure
\ref{figure:projections}.
If the inner flattening in the radial surface brightness profiles of Figure
\ref{figure:projections} were due to the presence of a core we would expect no
corresponding signature in the radial colour profile, since the scouring action of a
SMBH binary would not preferentially remove stars of a larger/smaller mass.
On the contrary, in Figure \ref{figure:cmap} we observe --- for both galaxies ---
a well defined ``bump'' corresponding to the physical extent of the rings
(compare with central panels of Figure \ref{figure:fit_models}).
For NGC~4494 we further observe that the innermost $\sim$20~pc are significantly
bluer than the outer profile: this feature is compatible with the emission by an
accreting disk and hence supports our previous statement about the presence of a
faint AGN.

Finally, we wish to stress that the dust rings described in this work
shall \emph{not} be confused with the torus surrounding the AGN
(although they are plausibly the large-scale source of material for the
accretion disk).
We show this by demonstrating that the inner radius of a torus is way smaller
than what we observe in our dust rings.
The inner radius of an AGN torus ($r_{sub}$) is dictated by dust sublimation, and
can be calculated as \citep{nenkova:2008}:

\vspace{-0.5cm}
\begin{eqnarray}
 r_{sub} \sim 0.4 ~ \left({L \over 10^{45}~erg~s^{-1}}\right)^{1/2} \left({1500~K \over T_{sub}}\right)^{2.6} & pc
 \label{equation:sublimation}
\end{eqnarray}

\noindent
where $L$ is the AGN luminosity and $T_{sub}$ is the dust sublimation temperature.
By using a luminosity of $\sim$10$^{39}$~erg~s$^{-1}$ (as is the case for the LLAGNs
in our galaxies) and adopting the standard $T_{sub} = 1500~K$, we obtain
$r_{sub} \sim 0.001$~pc.
This size is at least 2 orders of magnitude lower than what we measure for our
ring holes (Table \ref{table:fit}).
This simple test shows how the rings we observed are structures much larger
than an AGN torus.

\subsection{Evolution of dust rings and LLAGNs in ETGs}
\label{Evolution of dust rings and LLAGNs in ETGs}

\noindent
Dust structures are very common in ETGs \citep[e.g.][]{tran:2001}: their
morphology ranges from large-scale disks (such as in the famous case of the Sombrero
galaxy), to filaments, to inner disk rings.
Numerical simulations have shown that molecular disks result quite naturally from
wet mergers \citep[e.g.][]{xu:2010}, although recent integral field spectroscopy
results have shown that ETGs do not
necessarily require to accrete their dust content from merger events
\cite[e.g.][]{bassett:2017}.
These studies mostly concern large-scale structures, while the cases of inner rings
similar to those presented in this paper remain still mysterious.

\cite{lauer:2005} interpreted the compilation of dust phenomenology in their sample
of ETGs in evolutionary terms: they argued that the fate of dust structures is
inevitably to in-fall and settle at the center of the galaxy (independently of
their origins) in `episodic' events.
They support this theory with the observation that galaxies hosting well defined
inner rings are elsewhere devoid of dust.

With no intention of generalizing (given the limited sample analysed in this work)
we discuss here whether this scenario might be plausible for our objects;
NGC~4552 shows clear signs of disturbance in its extended light profile --- in fact,
it hosts a faint shell and a stellar stream \citep[e.g.][]{schweizer:1992} --- while
NGC~4494 does not present any evident `fine structure'.
By relating the fine structure to the time elapsed form the last merger
\citep[e.g][]{bonfini:sigma_core}, we can hence infer that NGC~4552 is dynamically
`younger' than NGC~4494, or, in other words, that the potential of the latter is more
relaxed than that of the former.
This observation, combined with the fact that NGC~4552 still hosts dust lanes and
that its ring is less defined than that of NGC~4494 (Figure \ref{figure:fit_models}),
might imply that dust settling is strictly related to the overall evolution
of the gravitational potential.

One representative comparison for our galaxies is NGC~4261, hosting the
prototypical, and arguably most studied dust disk in an ETG, first observed by
\cite{jaffe:NGC4261}.
These authors reported that the disk has a major-axis extent of 1$\arcsec$.79,
which at the distance of NGC~4261 \citep[29.4~Mpc;][]{jensen:NGC4261} corresponds
to a physical size of $\sim$250~pc, a thickness $\lesssim$20~pc, and an opening
angle of $\sim$10$^\circ$.
The galaxy has a prolate stellar geometry \citep{davies:NGC4261}, and displays
boxy isophotes \citep[][]{schweizer:1992}, but otherwise shows no sign of a past
merger, although the spatial distribution of its globular clusters hints toward
a recent fly-by interaction \citep[][]{bonfini:NGC4261}.
The dust geometry in NGC~4261 is clearly larger than the
rings observed in our sample galaxies.
However, the similar opening angle/thinness of the respective disks, and the
similarly relaxed potential of the galaxies suggest that the structure in NGC~4261
might be a scaled-up version closer to that observed in NGC~4494 rather than
that of NGC~4552.
The dust mass of the disk of NGC~4261 is estimated to be
log$(M_{dust}/M_{\odot}) = 4.7$ \citep{ferrarese:NGC4261}
while its AGN has a luminosity log$(L_{X} [erg~s^{-1}]) = 41.1$ in the 
2--10~keV band \citep{zezas:NGC4261}.
Hence, both its dust mass and X-ray luminosity are about 2 orders of magnitude
larger than for our sample galaxies: this hints to a direct connection between
the dust mass contained in the ring and the activity of the AGN. 

As a final remark we observe that, despite the two galaxies show rings with
similar mass and host LLAGNs of similar luminosity, their geometries
are very different in size. The most striking difference is that the ring hole in 
NGC~4552 is much smaller than that of NGC~4494.
This might be related to a different evolutionary state.
However, in order to further explore and constrain this connection, the methodology
presented in this work shall be applied to the study of a larger sample.

\section{Summary and conclusions}
\label{Summary and conclusions}

\noindent
Many ETGs exhibit a variety of dust features, and most notably inner ring structures
(Figure \ref{figure:fit_models}).
From an observational point of view, dust ring extinction can deeply modify a
galaxian surface brightness distribution.
To study the properties of the dust and its effects on the imaging
we created `Lord Of The Rings' (LOTR),
a pipeline which performs 2D fitting to galaxy images using surface
brightness models which account for the effects of dust (both
absorption and scattering).
Our pipeline acts in two steps.
At the first stage, it uses the 2D surface brightness modelling code \GALFIT{}
to determine the bona-fide unabsorbed stellar distribution
(Section \ref{Modelling of galaxian light profile}).
At the second stage, it uses the radiative transfer code \SKIRT{} to generate a
library of dust-absorbed stellar distributions sampling the dust mass and
dust geometry parameters.
These models are fit to the original image and the best-fit model is identified
via a simple \Chisq{} statistics (Section \ref{Monte-Carlo simulation of dust geometry and 2D fitting}).

We applied our methodology to NGC~4552 and NGC~4494, two ETGs reported to
host LLAGNs, and initially chosen to be representative of the core and
\mbox{core-\emph{less}} class (respectively).
However, our analysis revealed that NGC~4552 does \emph{not} host a core
\emph{nor} shows photometric indications for a central AGN  --- at variance with
previous studies based on similar optical imaging.
Instead, it showed that dust-obscuration mimicked the presence of a core, while
the central point-like luminosity was simply motivated by the stellar emission
shining through the central hole of the dust disk (Figure \ref{figure:fit_models}).
The case for NGC~4494 is similar.
However, for this galaxy we found evidence for a faint
$\sim$1.5 $\times$ 10$^{40}$ erg~s$^{-1}$ (ACS/$F555W$ band) AGN lurking at the center of
the dust structure (Section \ref{Dust pretending to be a core or an AGN}).

Building on the hypothesis of \cite{lauer:2005},
we suggest that the asymmetries in the ring geometries are symptomatic of the
evolutionary stages of the gravitational potential of the host galaxy.
Additionally, we find hints that the dust mass contained in the ring (although
not necessarily an extension of the AGN accretion torus) is linearly related to
the AGN luminosity.
Finally, we suggest that the size of the inner hole in the dust ring may 
help to study the evolutionary stage of a LLAGN.
In particular, one can discern whether a LLAGN is intrinsically faint because:
1) the infalling dust is currently reaching the galaxy center and it is just
now turning on the AGN engine (small hole), or instead because 2) past feedback has
pushed away the surrounding material and hence the AGN is currently fading out
due to starvation (large hole). 

An application of our methodology to a large collection of dust ring galaxies
(such as those reported by \citealt{lauer:2005} or  \citealt{tran:2001}) will
yield the first coherent parametrisation of central dust structures, and hence
provide important constraints on galaxy evolution.
More generally, our methodology can be used to `expose': 1) dust-obscured objects
disguised as core galaxies, and 2) stellar emission (shining through dust ring
holes) disguised as a point-like AGN.


\section*{Acknowledgements}

\noindent
We wish to thank the referee for the thorough review of this manuscript and
for the useful comments which helped us strengthening our conclusions. 
OGM wishes to acknowledge support by the UNAM PAPIIT grant (IA100516 PAPIIT/UNAM).
TB would like to acknowledge support from the CONACyT Research Fellowships.
GB acknowledges support for this work from UNAM through grant PAPIIT IG100115.
Based on observations made with the NASA/ESA Hubble Space Telescope, and obtained
from the Hubble Legacy Archive, which is a collaboration between the Space
Telescope Science Institute (STScI/NASA), the Space Telescope European
Coordinating Facility (ST-ECF/ESA) and the Canadian Astronomy Data Centre
(CADC/NRC/CSA).
This research has made use of the NASA/IPAC Extragalactic Database (NED),
which is operated by the Jet Propulsion Laboratory, California Institute of
Technology, under contract with the National Aeronautics and Space Administration.


\label{lastpage}

\end{document}